\documentclass{ws-procs10x7}
\usepackage{balance}
\usepackage{epsfig}
\newcolumntype{d}[1]{D{.}{.}{#1}}

\def\be{\begin{equation}}
\def\ee{\end{equation}}
\def\bea{\begin{eqnarray}}
\def\eea{\end{eqnarray}}

\def\Missing#1#2{{\mbox{$#1\kern-0.57em\raise0.19ex\hbox{/}_{#2}$}}}

\def\vMissing#1#2{\ifmmode
            \vec{#1}\kern-0.57em\raise.19ex\hbox{/}_{#2}
         \else
            {{\mbox{$\vec{#1}\kern-0.57em\raise.19ex\hbox{/}_{#2}$}}}
         \fi}
\def\lsim{\mathrel{\rlap{\lower4pt\hbox{\hskip1pt$\sim$}}
    \raise1pt\hbox{$<$}}}        
\def\gsim{\mathrel{\rlap{\lower4pt\hbox{\hskip1pt$\sim$}}
    \raise1pt\hbox{$>$}}}

\def\D0{D\O }

\def\simge{\mathrel{\rlap{\raise 0.53ex \hbox{$>$}}%
{\lower 0.53ex \hbox{$\sim$}}}}
\def\simle{\mathrel{\rlap{\raise 0.53ex \hbox{$<$}}%
{\lower 0.53ex \hbox{$\sim$}}}}

\makeindex
\begin{document}

\title{SEARCHES and PROSPECTS for the STANDARD MODEL HIGGS BOSON at the TEVATRON}

\author{Gregorio BERNARDI \\ on behalf of the CDF and D\O\ Collaborations}

\address{LPNHE, Universit\'es Paris VI et VII, France\\$^*$E-mail: gregorio@in2p3.fr}

\twocolumn[\maketitle\abstract{
We summarize the status of SM Higgs boson searches at the upgraded Fermilab Tevatron 
performed by the CDF and D\O\ collaborations, with an emphasis on measurements at
large Higgs mass, and derive sensitivity prospects 
for the upcoming increase in integrated luminosity.
}
]

\section{Introduction}

\begin{figure}[b]
\psfig{figure=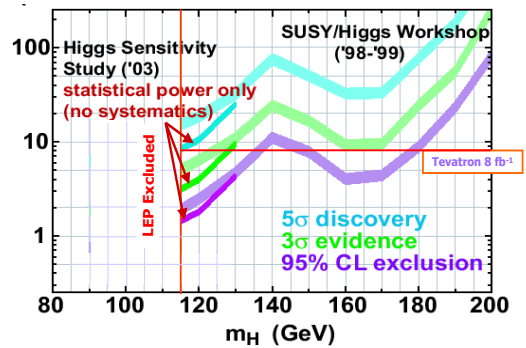,height=1.8in,width=2.8in,bbllx=83,bblly=235,bburx=550,bbury=510,clip=}
   \caption{
 Expected sensitivity to the Higgs boson at the Tevatron a function of $m_H$
and  luminosity.
\label{fig:ewwgSMH}}
\end{figure}

The Higgs boson is the only elementary scalar particle expected in the 
standard model (SM). Its discovery would be a major success for the SM and 
would provide new experimental insights into the electroweak symmetry breaking
mechanism. 
Direct measurements at LEP have excluded a SM Higgs boson with a mass 
$m_H < 114.4$ GeV at 95\% C.L. but
constraints from precision measurements  nevertheless favor a 
Higgs boson sufficiently light to be accessible at the Fermilab Tevatron 
Collider. 
The current preferred $m_H$ value, as deduced from a fit to 
electroweak measurements by the LEP, SLD, 
CDF, and D\O\ experiments~\cite{ewwg}
is  $85^{+39}_{-28}$ GeV. 
Combining with the direct limit,
yields a 95\% C.L. upper limit of 199 GeV.
At the Tevatron, indirect searches for the Higgs boson involve 
precision measurements of the masses of the top quark and $W$ boson, while 
direct searches require high luminosity samples for discovery or exclusion in 
the $115-185$ GeV mass range, as shown in Fig.~\ref{fig:ewwgSMH}.
At the Tevatron 
$p\overline{p}$ collider ($\sqrt{s}$=1.96~TeV), the 
two dominant mechanisms for Higgs production are 
gluon fusion, $gg \rightarrow H$, and associated production with a $W$ or $Z$ boson,
 $q\overline{q} \rightarrow W/Z + H$. 
The $gg$ fusion process has the largest cross section, $\sim$~1~pb at
$m_{H}$=115~GeV, 
but it is the most sensitive production mode only at relatively high mass,
$m_H >$135~GeV, where it has a dominant branching ratio in $WW^{*}$.
 At lower masses, the dominant decay $H \rightarrow b\bar{b}$ is 
swamped by multijet background, so 
 only the search for a Higgs produced in association
with a vector boson has sufficient sensitivity in this mass region.

Two simulation studies have been performed before the
"real" start of Run II to determine 
the sensitivity of the Tevatron experiments to SM Higgs physics.
The first study, in '98-99~\cite{susy-Higgs},
explored the whole mass range available with some approximation of the detector response,
while the second one (HSS, in '03) was restricted to the low mass region~\cite{Higgsens} and 
used a more realistic simulation, since the first data of Tevatron Run II had become available.
The second  study essentially confirmed the findings of the original study, and both
results are summarized in Fig.~\ref{fig:ewwgSMH}, after combination of all channels of both
experiments. The conclusions were that 
the estimated  integrated luminosity needed for a Higgs
discovery at $m_H =$ 115~GeV is approximately 8~fb$^{-1}$.  Evidence at 3 $\sigma$ might
be found with $\> 3-4 $fb$^{-1}$, while most of the Higgs mass region below
$\sim$185~GeV could be excluded at 95\% CL with $\sim 8$~fb$^{-1}$. 
The current results presented here below use about 10\% of the final luminosity which will 
be delivered.
We are thus now able to determine experimentally in which region the Tevatron is competitive with
the foreseen luminosity as  discussed at the end of this summary, after a brief
review of the machine performance, new Higgs search results  and their combinations.

\section{Tevatron Prospects}

The Higgs searches are crucially dependent on performance of the Tevatron accelerator and detectors,
and the detectors are currently performing close to their optimal design values.
The machine also is performing very well since the end of 2003, 
with close to designed delivered luminosities. As of today, about 2 fb$^{-1}$ have been delivered,
with weekly integrated luminosity routinely exceeding
20 pb$^{-1}$. %
\begin{figure}[htbp]
\psfig{figure=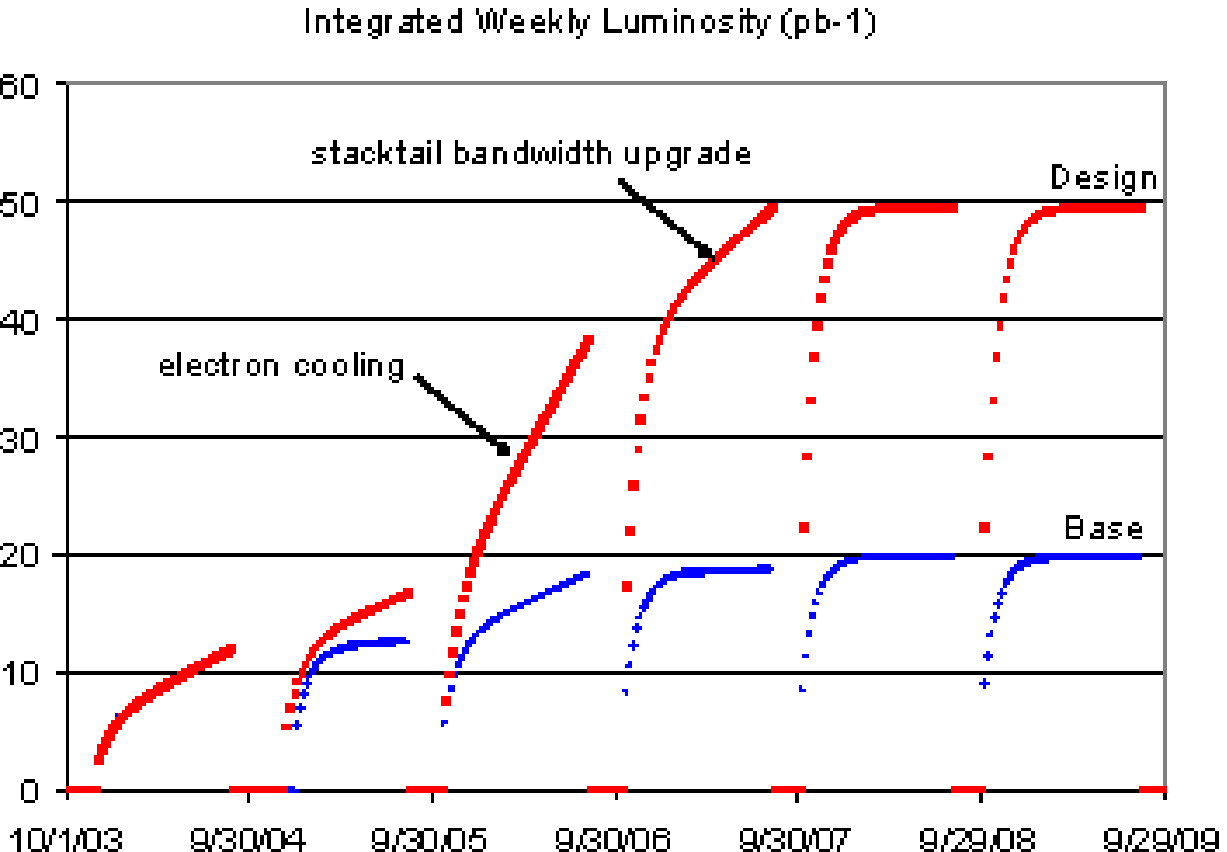,width=2.6in,height=1.8in
,bbllx=-10,bblly=-10,bburx=350,bbury=250,clip=} \\
\psfig{figure=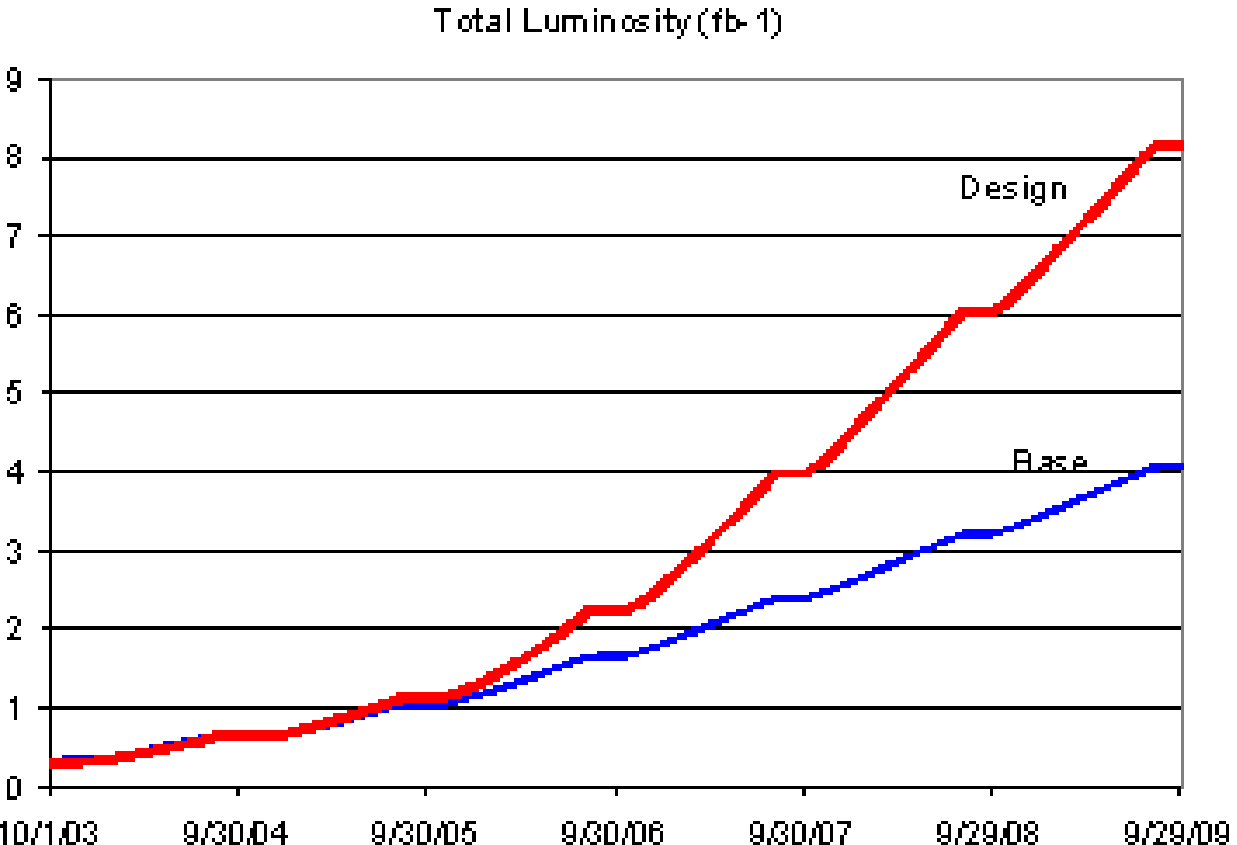,width=2.6in,height=1.8in
,bbllx=-10,bblly=-10,bburx=350,bbury=250,clip=}
  \caption{Weekly and total expected integrated luminosity of the Tevatron as
a function of time.
    \label{fig:tev} }
\end{figure}
Since this figure is the terminal value assumed in the minimal (so-called ``base'')
luminosity expectation (cf Fig.~\ref{fig:tev}a), the current performance ensure that
a minimal integrated luminosity of 4 fb$^{-1}$ will be achieved by
the end of 2009, as shown in Fig.~\ref{fig:tev}b. If the accelerator keeps following
the designed luminosity evolution, an integrated luminosity of about $8$ fb$^{-1}$ will be achieved,
rendering the potential for a Higgs discovery significant at the Tevatron. 

\section{Searches  for SM Higgs}

The searches at low mass in the $WH$ and $ZH$
production channels  have been presented in detail at this conference
in Ref.~\cite{ben}, so we concentrate
here on new results obtained assuming a large Higgs Mass
($> 135$ GeV). Two production channels have been exploited
so far: $p\bar{p}\rightarrow WH \rightarrow WWW^{\star}$, giving a final state
with 2 like-sign leptons,
and $p\bar{p}\rightarrow H \rightarrow WW^{\star}$ which gives a final state
with 2 opposite-sign leptons.
In the $WWW^{\star}$ channel, D\O\ has presented its published results on 
0.4 fb$^{-1}$~\cite{d0-www}:
after preselection, 34 like-sign events ($ee, e\mu, \mu\mu$) are left,
compared to a background of 34.9 events, dominated by instrumental
background, in particular by events in which one of the lepton charge
is misreconstructed. A topological likelihood discriminant based on 3
kinematic variables is then used to achieve further data reduction,
with a final number of observed vs. background events of 6 vs. 4.4.
In the absence of signal, the 95\% C.L. upper limit on 
$\sigma(WH) \times BR(H \rightarrow WW$) varies between 3.2 and 2.8 pb for $m_H$ between 
115 and 175 GeV. Although it is not the most sensitive channel in any
mass range, this channel 
provides additional sensitivity in the 130--150 GeV $m_H$ range.
There was low sensitivity
expected in this mass region (Fig.~\ref{fig:ewwgSMH}), but with this search,
the prospects in this area have now improved.

\begin{figure}[htbp]
\psfig{figure=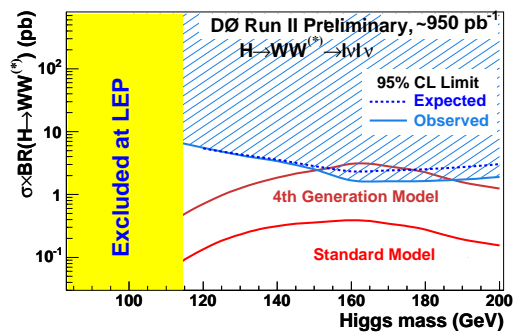,height=1.9in,bbllx=-10,bblly=-10,bburx=580,bbury=390,clip=}
  \caption{$WW^{\star}$  results at D\O\ with $\sim$1~fb$^{-1}$ 
of integrated luminosity, compared to the SM and to its extension with a 4$^{th}$
generation.
\label{wwd0}}
\end{figure}

In the and $ H \rightarrow WW^{\star}$ channel, CDF and D\O\ have 
already published results on samples of 0.35 fb$^{-1}$,
and obtained cross section limits of 3.5 pb$^{-1}$ at $m_H=160$~GeV\cite{d0-ww,cdf-ww}, 
where the SM expectation is 0.3 pb.
At this conference, D\O\ has presented updated results with $\sim$1~fb$^{-1}$
of data~\cite{d0-ww-em}. The search is similar to the published ones,
with a selection based on 2 isolated opposite-charge leptons $+$ missing transverse 
momentum, and further kinematic cuts to reduce the background, which is dominated
by $WW$ production. The number of events observed is 37, to be compared to
44.5 expected from SM background, and 1.7 for a SM Higgs with $m_H=$160 GeV.
The expected and observed limits are displayed in Fig.~\ref{wwd0}, and the limit
at 160 GeV has been reduced to 1.6 pb, i.e. only a factor
$\sim$5 away for the SM Higgs prediction. 
This result also excludes a SM with 4
families, having a Higgs with mass between 150 and 185 GeV, as shown. Indeed, the
quarks of the putative 4$^{th}$ family would enhance by a factor $~\sim$8
the Higgs production via the standard triangle diagram $ggH$~\cite{arik} in the
region of the search, assuming the most unfavorable case (infinite mass of the
4$^{th}$ generation quarks).

\section{Combined Limits on SM Higgs}

\begin{figure}[htbp]
\psfig{figure=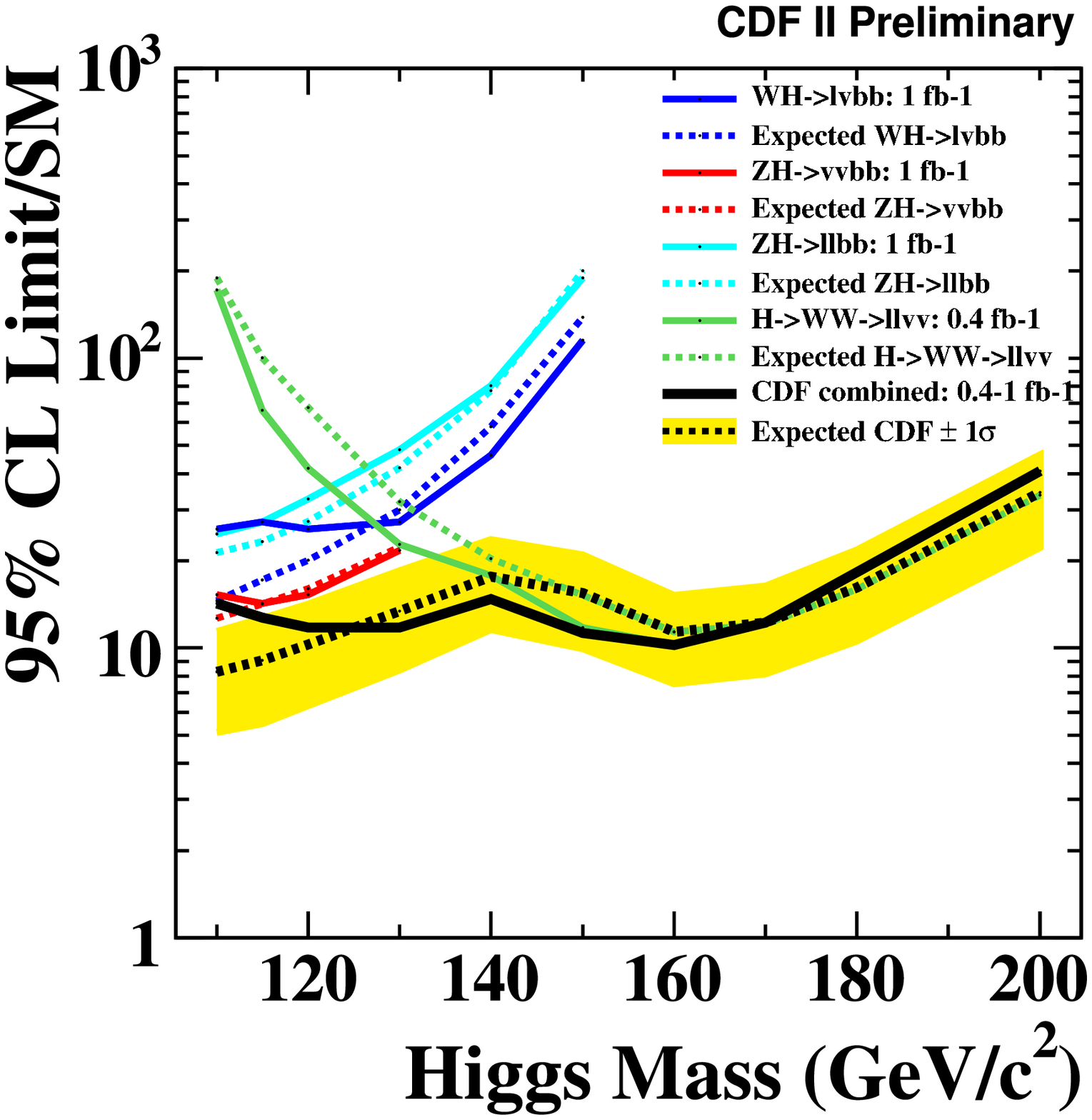,width=2.9in,height=2.1in}
\psfig{figure=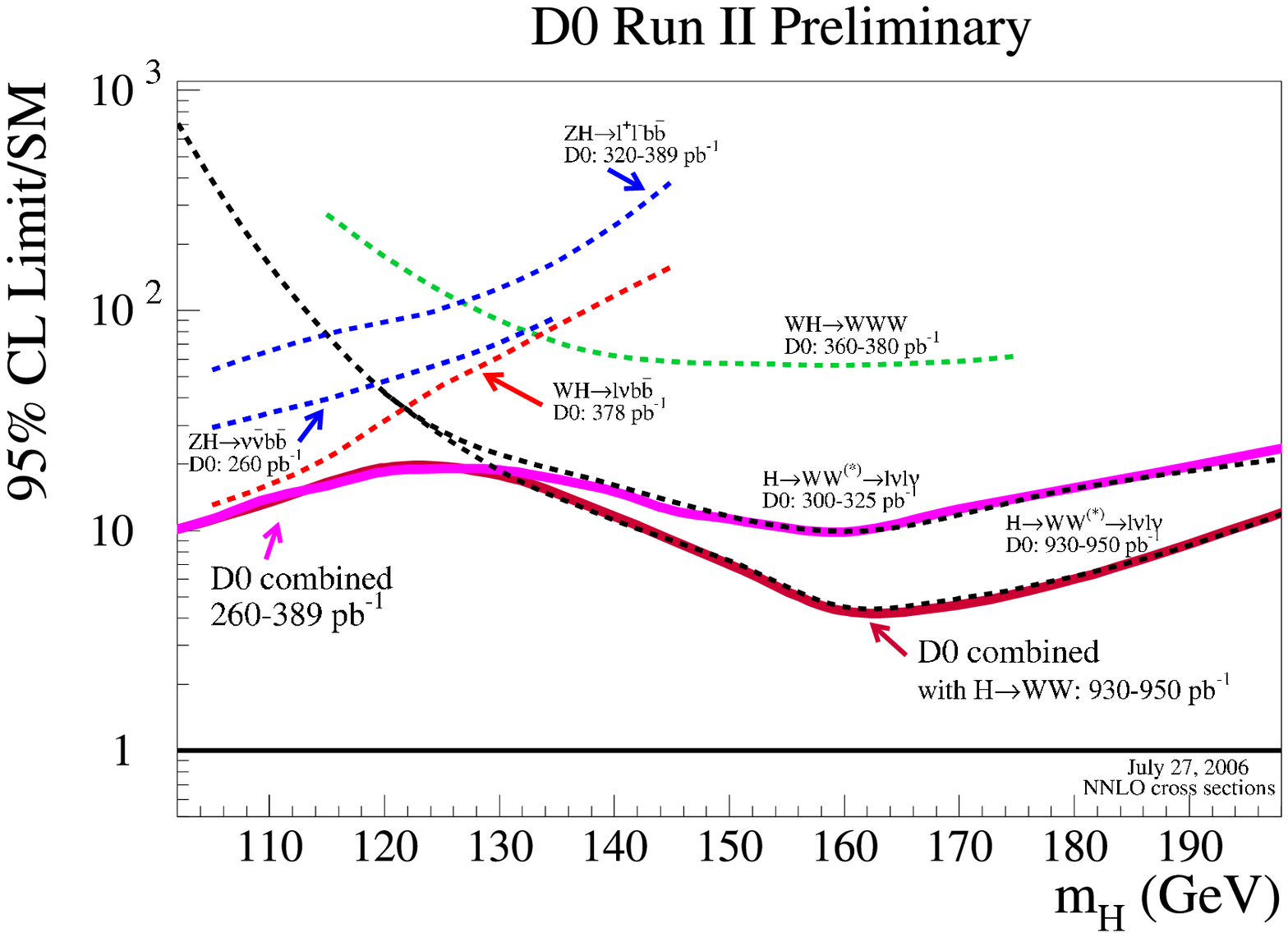,width=2.6in,height=2.0in,bbllx=30,bblly=240,bburx=520,bbury=620,clip=}
\caption{ 95\% C.L. limits on  Higgs production
cross section, divided by the SM expectation, as a function of $m_H$, for individual
channels and their combination of a) CDF and b) D\O .
A ratio of 1 is equivalent to a 95\% C.L. exclusion. 
\label{fig:wh-d0}}
\end{figure}

Both experiments have done the measurements in all channels, so the limits
can be improved by combining all channels into a single limit. 
To do that, CDF follows a Bayesian approach,
while D\O\ uses the $CL_s$ method developped for the Higgs search at LEP, see 
Ref.~\cite{cdf-d0} for details and complete references. 
The CDF (D\O ) results and their combinations are displayed in 
Fig.~\ref{fig:wh-d0}a(b). 

\begin{figure}[htbp]
\psfig{figure=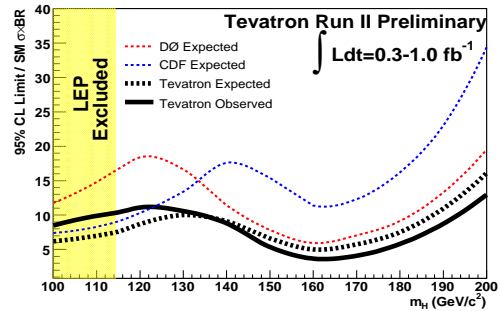,
height=1.7in,width=2.9in,bbllx=-10,bblly=-10,bburx=570,bbury=360,clip=}
\caption{CDF and D\O\ 95\% C.L. expected and 
observed combined limits on Higgs production
cross section, divided by the SM expectation, as a function of $m_H$.
The expected limits for CDF and D\O\ separately are also given
for comparison.
\label{combined}}
\end{figure}

Both combination methods have then been 
applied to the combination of the results of both experiments and the resulting
limits were found equal within 10\%.
The result of the CDF--D\O\ combination is shown in
Fig.~\ref{combined}. Also shown are the ``expected'' limits for the CDF--D\O\
combination, and for the CDF and D\O\ separate combination,  i.e. the limits
assuming that the observations would be exactly equal to the SM background
expectations. The expected limits of CDF and D\O\ have different shapes, since
presently CDF has analyzed the low Higgs mass channels
on a large dataset ($\sim$1 fb$^{-1}$)
and the high mass on a small dataset ($\sim$0.3 fb$^{-1}$, while D\O\ has done exactly
the reverse, so in fact the global curve corresponds to a good approximation
 to the curve of a single experiment with 1.3 fb$^{-1}$ for all channels.
The expected limits, which allow to judge the current sensitivity, show that at
115 (160) GeV, the limit is 7.6 (5.0) times higher than the SM expectation.
The increase to the expected design integrated luminosity will reduce these values
by approximately a factor 3.5, so a further gain in sensitivity of about 2.2 (1.4)
is needed in  data analysis to reach exclusion at these mass points.
Separate studies already performed in the two
experiments show that this gain will definitely be greater than 2.5 at 
low mass (use of Neural Net
techniques in selection and in $b$-identification, better dijet mass resolution,
increased acceptance, reduced systematic errors and inclusion of $\tau$ channels)
and 1.7 at high mass (where $b$-identification and dijet mass resolution do not
play a role). With further developments under test, or already used
in other analyses, the full mass range
between 115 and 185 GeV will be sensitive to an exclusion (or an evidence) at
the 95\% C.L. and at more than 3 $\sigma$ level in the most sensitive regions
($\sim$ 115 or 160 GeV).

In conclusion, the SM Higgs searches
 in all the channels are now being regularly updated
and combined. We expect that 
after implementing the analysis improvements mentioned above,
the sensitivity prospects will be met. 
Hence, the prospects for uncovering the first evidence of a light mass Higgs boson with  
4--8 pb$^{-1}$ of Tevatron integrated luminosity  are real.


\begin{thebibliography}{9}


\bibitem{ewwg} S.\ Eidelman {\it et al}, Phys.\ Lett. B {\bf 592}, 1 (2004); 
LEP Electroweak Working Group,\\
http://lepewwg.web.cern.ch/LEPEWWG/

\bibitem{susy-Higgs} M. Carena, J. Conway, H. Haber, J. Hobbs et al.,
 FERMILAB-CONF-00/279-T (2000), hep-ph/0010338.

\bibitem{Higgsens} CDF and D\O\ Collaborations, FERMILAB-PUB-03/320-E (2003).

\bibitem{ben}B. Kilminster, XXXIII International Conference on High Energy Physics, Moscow, 
2006, {\tt
http://tevnphwg.fnal.gov/results/ }

\bibitem{d0-www}  D\O\ Collaboration,
V.M.~Abazov {\it et al.},  Phys. Rev. Lett. 97, 151804 (2006). 
\bibitem{d0-ww}  D\O\ Collaboration,
V.M.~Abazov {\it et al.},  Phys. Rev. Lett. 96, 011801 (2006). 
\bibitem{cdf-ww}  CDF Collaboration,
A.~Abulencia {\it et al.},  Phys. Rev. Lett. 97, 081802 (2006).

\bibitem{d0-ww-em}  D\O\ Collaboration,
V.M.~Abazov {\it et al.}, \\
{\tt
http://www-d0.fnal.gov/Run2Physics/WWW/ \\
results/prelim/HIGGS/H16/H16.pdf \\
http://www-d0.fnal.gov/Run2Physics/WWW/ \\
results/prelim/HIGGS/H21/H21.pdf}

\bibitem{arik}E. Arik {\it et al.}, {\tt arXiv:hep-ph/0502050},\\
 E. Arik {\it et al.}, Eur. Phys. J. C 26, 9 (2002).

\bibitem{cdf-d0}TevNPH working group, for the CDF \ and D\O\ Collaborations,{\tt arXiv:hep-ex/0612044}.


\end{thebibliography}
\end{document}